\documentclass[10pt,letterpaper]{article}
\usepackage{opex3}
\usepackage{color}
\usepackage{threeparttable}
\usepackage{cite}
\usepackage{cancel}

\begin{document}

\title{Polarization-entangled photon-pair source obtained via type-II non-collinear SPDC process with PPKTP crystal}
\author{Sang Min Lee,$^1$ Heonoh Kim,$^1$ Myoungsik Cha,$^1$ and Han Seb Moon$^{1,*}$}
\address{$^1$Department of Physics, Pusan National University, Geumjeong-Gu, Busan 609-735, Korea}
\email{$^*$hsmoon@pusan.ac.kr}

\begin{abstract}
We demonstrate a polarization-entangled photon-pair source obtained via a type-II non-collinear quasi-phase-matched spontaneous parametric down-conversion process with a 10-mm periodically poled KTiOPO$_4$ crystal, which is as stable and wavelength-tunable as the well-known Sagnac configuration scheme. A brightness of 4.2 kHz/mW is detected and a concurrence of 0.975 is estimated using quantum state tomography. Without loss of entanglement and brightness, the photon-pair wavelengths are tunable through control of the crystal temperature. This improvement is achieved using the non-collinear configuration and a stable interferometric distinguishability compensator.
\end{abstract}


\section{Introduction}
Many significant experiments concerning topics related to quantum information processing, such as quantum teleportation~\cite{Bouwmeester}, dense coding~\cite{Mattle}, quantum key distribution~\cite{Poppe}, and controlled-not gate~\cite{Gasparoni}, have been conducted using polarization-entangled photon pairs generated via spontaneous parametric down-conversion (SPDC) processes with $\beta$-barium borate (BBO) crystals \cite{Kwiat1,Kwiat2,Giorgi,KimY}. However, these sources have relatively low ($\leq$ 140 Hz/mW) brightness ($BR$; coincidence counts per second per pump power, without loss-correction) because of inefficient collection of the generated photons and small nonlinear coefficient. To date, considerable efforts have been made towards achieving greater brightness and higher entanglement quality in such cases. In particular, in the early 2000s, nonlinear crystals with periodically poled (PP) structures and quasi-phase-matching (QPM) conditions yielded remarkably increased brightness~\cite{Tanzilli,Banaszek,Sanaka} without polarization entanglement. To realize polarization-entangled photon-pair sources with QPM crystals, schemes involving post-selection~\cite{Kuklewicz}, crossed crystals~\cite{Pelton, Hubel, Steinlechner12}, a folded Mach-Zehnder interferometer (MZI)~\cite{Fiorentino, Konig}, a polarization Sagnac interferometer (PSI)~\cite{KimT, Wong, Fedrizzi}, two crystals Sagnac interferometer~\cite{Stuart}, and linear double-pass geometry~\cite{Steinlechner13} have been demonstrated, which are applied under collinear conditions. In particular, two outstanding schemes have been reported by Fedrizzi {\it et al}.~\cite{Fedrizzi} and Steinlechner {\it et al}.~\cite{Steinlechner13}, who detected 271- and 390-kHz/mW/nm spectral brightness ($SB$; $BR$ per bandwidth) and more than 99\% fidelity with one of the Bell states using type-II and type-0 SPDC processes with a PP potassium titanyl phosphate (KTiOPO$_4$; PPKTP) crystal, respectively, with stable configurations. However, the schemes reported in those studies involve an intricate alignment or specialized optics such as a dual-wave plate, a specifically tailored wave plate, and a temperature-stabilized compensation crystal.

Here, we demonstrate a polarization-entangled photon-pair generation scheme~\cite{Jeong1} based on a type-II non-collinear PPKTP crystal and employing a continuous-wave (cw) pump laser~\cite{Fiorentino} and an MZI-configured distinguishability compensator~\cite{KimY}. Our source is  stable and relatively easy to build up because of the simple configuration and lack of specialized optics. The pump-power $P_p$ dependence of the source is examined in order to estimate the $BR$ and pair-production-rate (loss-corrected $BR$), and the crystal-temperature $T$ dependence is monitored in order to demonstrate wavelength ($\lambda$) tunability. We also conduct quantum state tomography (QST) \cite{James} to evaluate the entanglement quality of the generated states. The detected $BR$, $\lambda$ tunability, and concurrence (one of proper measures of the degree of entanglement~\cite{Wootters}) are 4.2 kHz/mW, 0.198 nm/$^\circ$C, and 0.975, respectively. We expect that enhanced $BR$, $SB$ and a greater degree of entanglement can be achieved if optimized focusing parameters, a longer crystal, and a shorter coincidence window are employed.

The remainder of this paper is organized as follows. In section~\ref{s2}, we introduce and explain the experiment setup of our polarization-entanglement photon-pair source and measurement apparatus using a schematic diagram. In section~\ref{s3}, we briefly describe the QPM condition of the type-II non-collinear SPDC process and the theoretical foundation behind the expected pair-production-rate through comparison with a type-II collinear PSI setup. We also provide the results of numerical calculations, which are used as a reference for comparison against the experimental results presented in section~\ref{s4}. From those experimental results, we estimate and discuss the $BR$, pair-production-rate, $\lambda$ tunability, entanglement quality, and stability of our source. Finally, a summary is presented in section~\ref{s5}.

\begin{figure}[t]
\centerline{\includegraphics[scale=0.45]{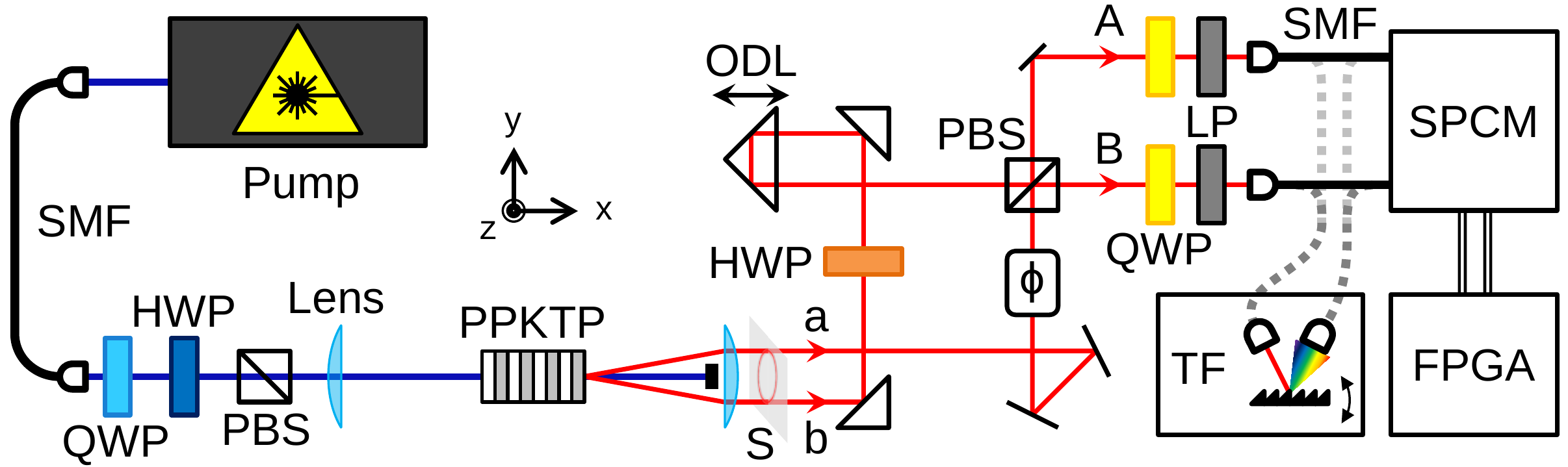}}
\caption{Schematic diagram of experimental setup. The pump is a single-longitudinal-mode cw laser with a center wavelength $\lambda_p$ of $\sim$406.2 nm and a full-width at half-maximum (FWHM) bandwidth of $\Delta\nu$ = 0.2(1) GHz. SMF: Single-mode fiber for 405 or 780 nm; QWP (HWP): Zero-order quarter (half)-wave plate for 405 or 808 nm; PBS: Broadband polarizing beam-splitter; Lens: Anti-reflection (AR)-coated plano-convex lens (focal length $f$ = 200 mm, coating designation: A or B); PPKTP: periodically poled KTiOPO$_4$ ($\Lambda$ = 10 $\mu$m; length $L$ = 10 mm); S: temporary image screen to measure photon rings; ODL: Optical delay line consisting of right-angled prism and $x$-axis translation stage; $\phi$: Phase shifter consisting of sequential wave plates (QWP-HWP-QWP); LP: Broadband linear polarizer; TF: Temporally used wavelength tunable filter (FWHM of $\Delta \lambda$ = 1.82 nm); SPCM: Single-photon counting module; FPGA: Single and coincidence counting unit.}
\label{setup}
\end{figure}

\section{Scheme and methods} \label{s2}
Figure~\ref{setup} is a schematic diagram of our experimental setup for the polarization-entangled photon-pair source. A single-longitudinal-mode cw laser 
is employed as a pump, having a center wavelength $\lambda_p$ and full-width at half-maximum (FWHM) bandwidth $\Delta \nu$  measured at approximately 406.2 nm and 0.2 GHz, respectively; these measurements were obtained using a spectrometer ({\it{Ocean Optics}}, USB4000; $\Delta \lambda$ $\simeq$ 1 nm) and a scanning Fabry-Perot interferometer ({\it{Thorlabs}}, SA210; $\Delta \nu$ = 67 MHz), respectively. The pump beam is spatial-mode-filtered by a single-mode fiber (SMF) for 405 nm and its polarization and $P_p$ are adjusted by wave plates and a polarizing beam-splitter (PBS). The well-defined pump beam is focused by a lens (focal length $f$ = 200 mm, ARC-A) and passes through the PPKTP crystal ($\Lambda$ = 10 $\mu$m, $L$ = 10 mm) in the Rayleigh range. Under the type-II non-collinear SPDC condition, a pump-light photon is probabilistically converted to a pair of photons with horizontal ($y$-axis; $H$) and vertical ($z$-axis; $V$) polarization and spatial or directional modes determined by the phase-matching (PM) condition. In our setup, we select two spatial modes ($a$ and $b$) that are tilted at approximately 0.85$^\circ$ relative to the $x$-axis, cento-symmetric to each other, and along the $xy$-plane. The two spatial modes are collimated by a lens ($f$ = 200 mm, ARC-B). Each photon ($H$ or $V$ polarized) of a pair can be emitted in either of the two modes ($a$ or $b$), but the counterpart photon must then be emitted  in the other direction (i.e., $b$ or $a$, respectively),  in accordance with the law of conservation of momentum for near-degenerate cases ($\lambda_H \simeq \lambda_V$, where $\lambda_H$ and $\lambda_V$ are the center wavelengths of the $H$ and $V$ polarized photons, respectively). Therefore, if all other degrees of freedom are indistinguishable, the polarizations of a photon pair in modes $a$ and $b$ are entangled as $|HV\rangle_{ab}+|VH\rangle_{ab}$. However, the wavelengths can be distinguishable if the difference between $\lambda_H$ and $\lambda_V$ in a given mode is significantly larger than the wavelength bandwidth $\Delta \lambda$ of each photons. Furthermore, in general, the refractive indices of the $H$ and $V$ polarizations in the crystal differ, so photons of two polarizations in a given mode are temporally distinguishable because of the group velocity difference; this holds even for the degenerate case. Thus, the generated polarization state is entangled with other degrees of freedom as $|H(\lambda_H,t_H)\rangle_a |V(\lambda_V,t_V)\rangle_b + |V(\lambda_V,t_V)\rangle_a |H(\lambda_H,t_H)\rangle_b$. To compensate for this distinguishability, we adopt an additional setup~\cite{KimY, Fiorentino, Jeong1} which consists of a zero-order half-wave plate (HWP) at 45$^\circ$, an optical delay line (ODL), and a broadband PBS. After proper adjustment of the ODL, the polarization states of a photon-pair in modes $A$ and $B$ of Fig.~\ref{setup} are entangled as
\begin{equation}
\label{phi}
|\psi \rangle _{AB} = \frac{|HH\rangle_{AB}+e^{i \phi}|VV\rangle_{AB}}{\sqrt{2}} \otimes |\lambda_A \lambda_B \rangle_{AB},
\end{equation}
where $\lambda_A$ and $\lambda_B$ are identical to $\lambda_H$ and $\lambda_V$ in modes $a$ or $b$, respectively. In Eq.~(\ref{phi}), the entire state of the wavelengths for modes $A$ and $B$ is separable from the polarization state, and the time information depending on the polarizations is negligible. We neglect the entanglement between the $\lambda$ for modes $A$ and $B$ in Eq.~(\ref{phi}), as the $\Delta \lambda$ of each photon is sufficiently narrow for non-degenerate conditions~\cite{Fedrizzi, Ramelow} and the spectral correlation can be erased using narrow interference filter (IF) or engineered by a group-velocity matching condition with a broadband pump~\cite{Grice, Keller, Mosley, Jin}.

A relative phase $\phi$ between the two states in Eq.~(\ref{phi}) can be introduced through unexpected birefringence of the optics in the interferometric compensator; therefore, we insert a phase shifter consisting of sequential zero-order wave plates (QWP-HWP-QWP; quarter-wave plate = QWP) to adjust $\phi$. The fast axes of the QWP are fixed at 45$^\circ$ from the horizontal direction, and we adjust the HWP angle to obtain $\phi = 0$.

To measure the polarizations of the generated state, we introduce polarization analyzers consisting of a QWP and a broadband linear polarizer (LP). After passing the analyzers, the photons are coupled into SMFs for 780 nm via aspheric lenses ($f_e$ = 8 mm) and detected using single-photon counting modules (SPCM). Electric pulses from the SPCM are counted by a field-programmable gate array (FPGA) system calibrated for single and coincidence counts (coincidence window $\simeq$ 55 ns). We also measure the photon $\lambda$ values in modes $A$ and $B$ using a wavelength tunable filter (TF, FWHM = 1.82 nm). In Fig.~\ref{setup}, the temporary image screen labeled $S$ consists of an LP, an IF (three-cavity bandpass filter at approximately 812 nm, FWHM = 3 nm), an aspheric lens ($f_e$ = 8 mm), and a multi-mode fiber (MMF), which are mounted on a $yz$-axis translation stage. By counting the number of photons coupled in the MMF during stage movement, we can record 1D or 2D photon ring images.

\section{Theory and numerical calculations} \label{s3}
A photon pair generated  in a type-II non-collinear SPDC process with a PPKTP crystal adheres to the QPM conditions \cite{Armstrong, Fejer}, such as: (i) energy conservation: $\omega_p = \omega_H + \omega_V$; and (ii) momentum conservation: $\vec k_p = \vec k_H + \vec k_V + \vec K$, where $\omega$ and $\vec k$ are the angular frequency and wavevector of the photons, respectively, the $p$ subscript denotes the pump, and $\vec K$ is the  grating vector of the periodic structure of the PPKTP crystal. If $\Lambda$ is the period of the poled pattern, $K$ is defined as $2\pi/\Lambda$. The coordinate system depicted in Fig.~\ref{phi} corresponds to the principal axes of the PPKTP crystal, and $\vec k_p$ and $\vec K$ are parallel to the $x$-axis. In the course of the type-II SPDC process, the nonlinear coefficient $d_{24}$ is utilized, which means that the pump is $H$ polarized (in the $y$-direction) and the two generated photons are $H$ and $V$ polarized (in the $y$- and $z$-directions, respectively). See \cite{Lee} for a more detailed description of the QPM condition of PPKTP.

In the collinear case, QPM condition (ii) is reduced to $k_p = k_H + k_V + K$, and the refractive indices of the $H$ and $V$ photons are functions of the wavelength and $T$ of the crystal. KTP is a biaxial crystal and its refractive indices for principal axes depending on wavelength and $T$ have been well studied~\cite{Fan, Bierlein, Wiechmann, Kato, Emanueli}. Here, we adopted the $T$-dependent Sellmeier formulas and thermal expansion coefficients for KTP from \cite{Fiorentino, Kato, Emanueli}. Using numerical calculations with given parameters, such as $\lambda_p$ = 406.2 nm and $\Lambda$ = 10 $\mu$m, we obtained the $T$ value of the degenerate collinear case as $T_{dc}$ = 98.98$^\circ$C. We should note that numerically calculated values of $T_{dc}$ differ depending on the references used for the  Sellmeier formulas, as these expressions are empirical relationships obtained from individual finite experimental datasets. Therefore, the following numerical results for the non-collinear conditions are not ideal references for comparison against the experimental results. Instead, they provide a rough outline only.

Under general non-collinear SPDC in biaxial crystals, accurate prediction of PM conditions in the entire space is sophisticated~\cite{Ito}. However, it is much easier to calculate the PM condition only on the two orthogonal principal planes ($xy$- and $xz$-planes), as the limiting cases. Furthermore, for a near-collinear case where the angle between the wave-vectors of the pump and the SPDC is small (3.0$^\circ$ or smaller as in our experiment), one can neglect the Poynting vector walk-off from the wave-vector (0.3$^\circ$ or less), significantly simplifying the calculation process.

Under near-collinear conditions, the ring images of the $H$ and $V$ photons are overlapped and indistinguishable in the degenerate case, and the effective nonlinear coefficient is also almost identical to that of the collinear case ($d_{eff} = 2 d_{24} / \pi$ in the ideal case). At this point, we note that photon pairs are accumulated in two spatial modes using the non-collinear setup, so that twice the number of photon pairs can be collected compared with the collinear setup. This means that our scheme, which is based on non-collinear conditions, can yield the twice brightness of the collinear configuration in an ideal experimental setup, i.e., with the same $L$ and optimized spatial mode parameters for pump and generated photons.

\begin{figure}[b]
\centerline{\includegraphics[scale=1]{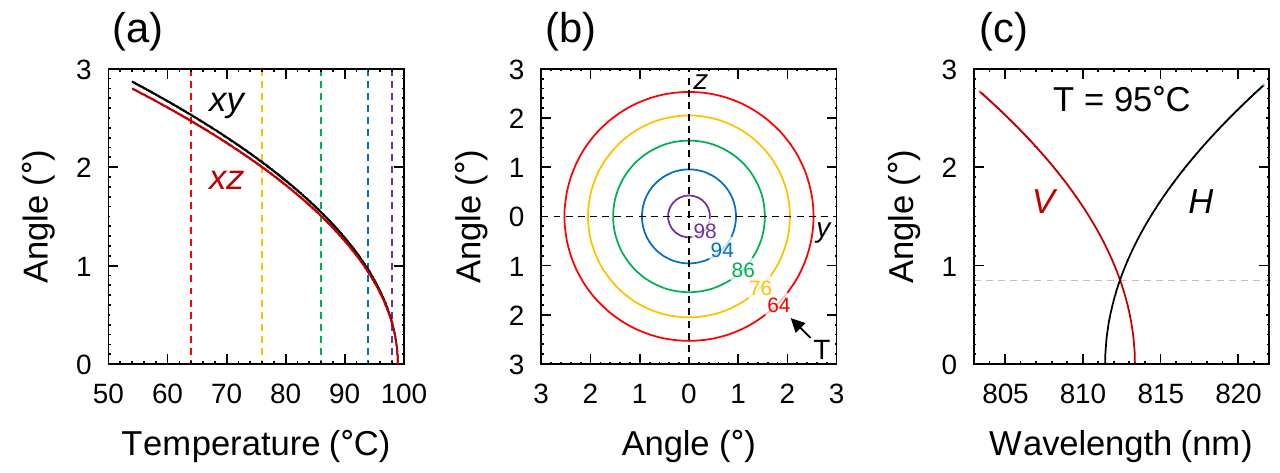}}
\caption{Numerically calculated emission angles $\theta$ for photons emitted from PPKTP crystal exit plane for: (a) Degenerate case on $xy$- and $xz$-planes as a function of $T$; (b) degenerate case for several fixed $T$ values on $yz$-plane; and (c) fixed $T$ = 95$^\circ$C on $xy$-plane as a function of photon wavelength.}
\label{cal}
\end{figure}

\begin{figure}[t]
\centerline{\includegraphics[scale=1]{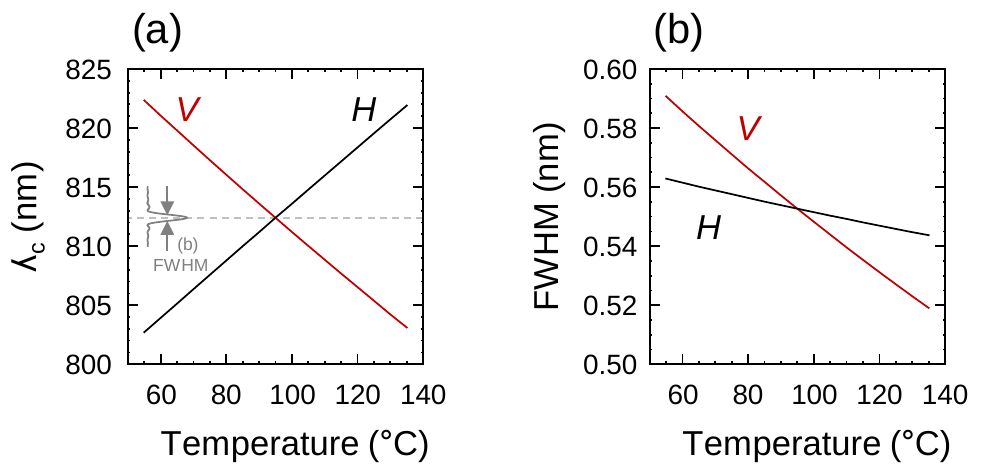}}
\caption{$T$ dependence of (a) $\lambda_H$ and $\lambda_V$ and (b) FWHM $\Delta \lambda$ of $H$ and  $V$ photons of pair generated under non-collinear condition, with spatial modes determined by degenerate condition at $T$ = 95$^\circ$C.}
\label{tune}
\end{figure}

Figure~\ref{cal} shows the approximated (assuming near-collinearity) numerical results for: (a) The emission angle ($\theta$) values outside the crystal exit plane of the degenerate  ($\lambda_H$ = $\lambda_V$) SPDC cone as a function of $T$ for the photon pairs emitted on the $xy$- and $xz$-planes; (b) ring (or cone) images on the $yz$-plane at several fixed $T$ values for the degenerate case; and (c) the $\theta$ values on the $xy$-plane according to the wavelengths of the $H$ and $V$ photons at a fixed $T$ = 95$^\circ$C. As shown in Fig.~\ref{cal}(a), the $\theta$ values of the degenerate photon pairs on the $xy$- and $xz$-planes for fixed $T$ differ slightly, because of the difference between the principal indices $n_y$ ($\sim n_x$) and $n_z$~\cite{elliptic}. Therefore, the ring images of the photon pairs in Fig.~\ref{cal}(b) are slightly elliptic ($e$ $\simeq$ 0.22)~\cite{Fiorentino}. The specific value of $T$ = 95$^\circ$C for Fig.~\ref{cal}(c) was chosen to satisfy the fact that the $\theta$ value for the degenerate case is approximately 0.85$^\circ$, which is the real experimental parameter of our setup. At the spatial modes ($\theta\simeq0.85$, the degenerate condition for $T$ = 95$^\circ$), the derivative of the emission angle against the wavelength $d\theta/d\lambda$ is approximately 0.45 $^\circ$/nm. This is a characteristic of the non-collinear setup that contrasts with the collinear configuration, as $d\theta/d\lambda$ for the degenerate collinear condition ($\theta = 0$) is approximately infinite.

Figure \ref{tune} shows the numerical results for the (a) $\lambda_H$ and $\lambda_V$ and (b) spectrum FWHM $\Delta \lambda$ of the $H$ and $V$ photons as a function of $T$, where the spatial modes were fixed to the degenerate condition in Fig.~\ref{cal}(c). In fact, for non-degenerate cases ($\lambda_H \neq \lambda_V$), the $\theta$ values of the $H$ and $V$ photons differ slightly, in accordance with the law of conservation of momentum. However, this difference is negligible, as $\Delta \theta_{HV} / \Delta \lambda_{HV}$ $\simeq$ 18 $\mu$rad/nm, where $\Delta \theta_{HV}$ and $\Delta \lambda_{HV}$ are the differences in the emission angles and wavelengths between the $H$ and $V$ photons of a given pair, respectively. $\lambda$ tunability via control of $T$ is shown by the slopes of the trendlines in Fig.~\ref{tune}(a), which were calculated as 0.23 nm/$^\circ$C. The numerical results in Fig.~\ref{tune}(b) demonstrate that the photon $\Delta \lambda$  decrease slightly as $T$ increases, and the degenerate case ($\lambda_H$ = $\lambda_V$ = 812.4 nm) is given by 0.553 nm, which is almost the same result as that obtained for the collinear case \cite{Fedrizzi}.

\section{Experimental results and discussions} \label{s4}
\subsection{Photon ring images}

\begin{figure}[b]
\centerline{\includegraphics[scale=0.45]{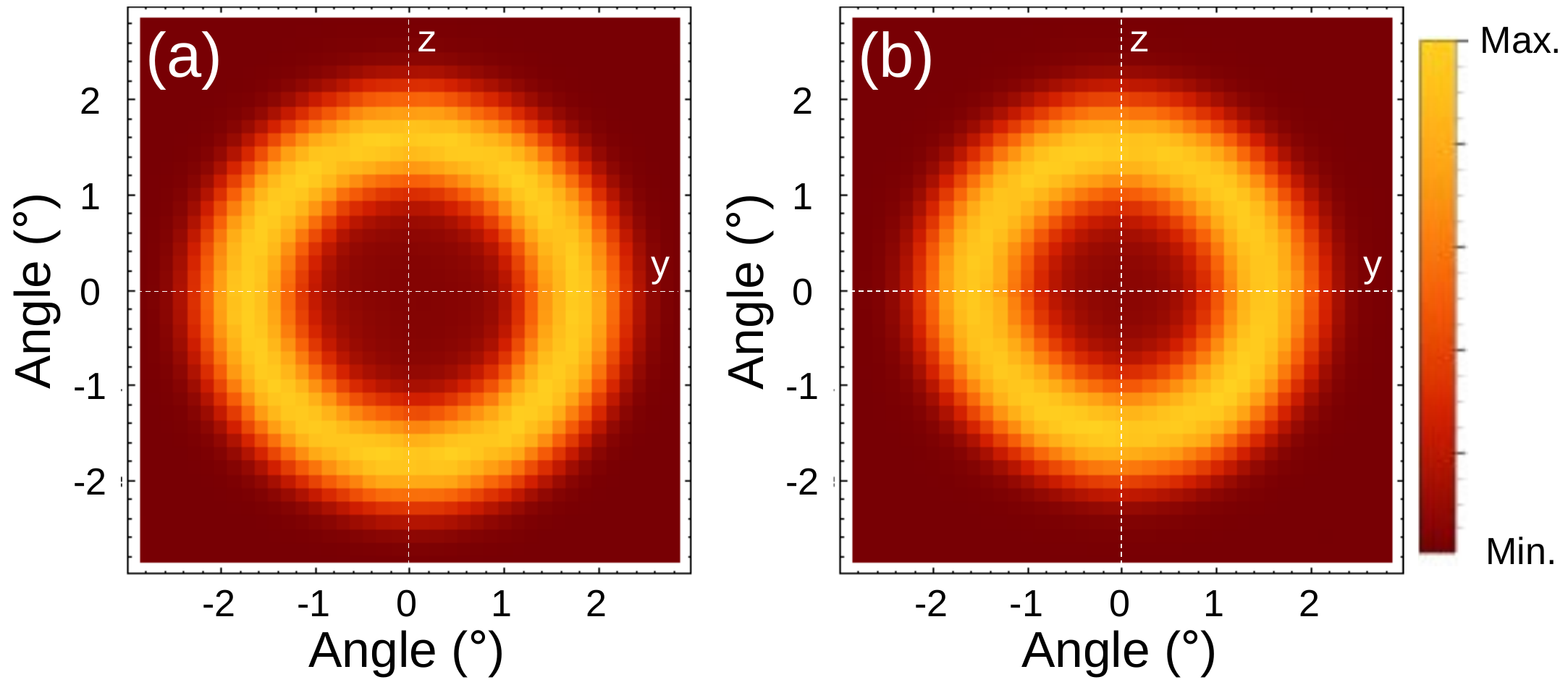}}
\caption{Measured photon ring images for (a) $H$ and (b) $V$ polarized photons with IF (IF center wavelength $\lambda_{IF}$ $\simeq$ 812 nm, FWHM = 3 nm).}
\label{ring}
\end{figure}

Firstly, to mark the center of the ring images or to determine the position of the collinear condition, we measured the photon ring images at $T$ = 68.6$^\circ$C, as shown in Fig.~\ref{ring}, using the $S$ shown in Fig.~\ref{setup}. To show the ring images clearly, we decreased $T$ to a value significantly lower than the calculated $T_{dc}$ = 98.98$^\circ$C, as the IF $\Delta \lambda$ (FWHM = 3 nm) was broader than the numerically calculated photon $\Delta \lambda$ ($<$ 0.6 nm) in 10-mm PPKTP. Figures \ref{ring}(a) and \ref{ring}(b) show ring images for $H$ and $V$ polarized photons, respectively. The measured ring image in Fig.~\ref{ring}(a) is slightly larger than that of Fig.~\ref{ring}(b), because the IF center wavelength $\lambda_{IF}$ may not have been identical to the degenerate wavelength (2$\lambda_p$). By means of data fitting, the ring-image ellipticities were estimated to be 0.25(1).

\begin{figure}[t]
\centerline{\includegraphics[scale=0.45]{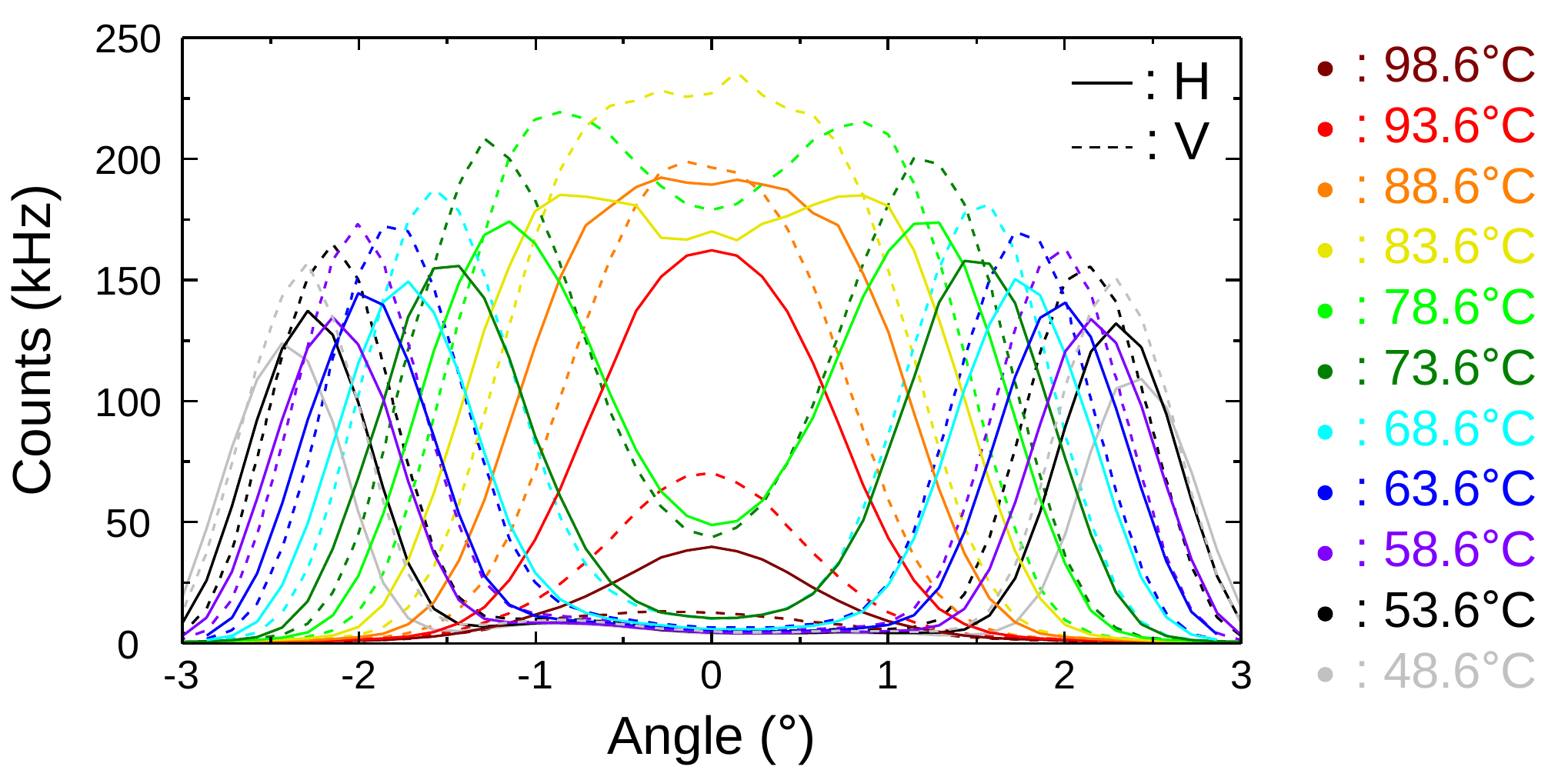}}
\caption{Measured cross-section (1D) images of photon rings along $y$-axis for various $T$ (48.6 - 98.6 $^\circ$C) with IF (FWHM = 3 nm).}
\label{ycross}
\end{figure}

Figure \ref{ycross} shows experimentally observed cross-section images of photon rings along the $y$-axis for $T$ varied from 48.6 to 98.6$^\circ$C. Because of the broad $\Delta \lambda$ of the IF and the discrepancy that $2\lambda_p\neq\lambda_{IF}$, we could not accurately determine the experimental value of $T_{dc}$ by means of the cross-section results shown in Fig.~\ref{ycross}. However, we roughly estimated that its value was approximately 89(5)$^\circ$C. It is possible to obtain an accurate value of $T$ for the degenerate conditions of the collinear and non-collinear cases by measuring the spectra of $H$ and $V$ photons using the TF. The spectral results for a non-collinear case are discussed below. In Fig.~\ref{ycross}, the maximum count rates of the $H$ photons are larger than those of the $V$ photons for $T$ = 98.6 and 93.6$^\circ$C, which are above the experimental $T_{dc}$ value of 89(5)$^\circ$C, and the angles of the peak counts for the $H$ photons are larger than those of the $V$ photons at values lower than the $T_{dc}$. From these results, we inferred that $\lambda_{IF}$ was slightly longer than the degenerate wavelength, i.e., $\lambda_{IF} - 2\lambda_p \simeq $  0.07 nm. Further, we noted that the peak counts of the $H$ and $V$ photons are reduced when $T$ decreases, as the near-collinear approximation may not work for lower $T$, and $d_{eff}$ is decreased. Therefore, we chose the SPDC cone angle approximately 0.85$^\circ$ with respect to the pump beam in order to maintain a similar counting rate in each spatial mode and to collect approximately twice the number of photon pairs compared to the collinear case.

\subsection{Temperature dependence of photon-spectrum}

\begin{figure}[h]
\centerline{\includegraphics[scale=0.45]{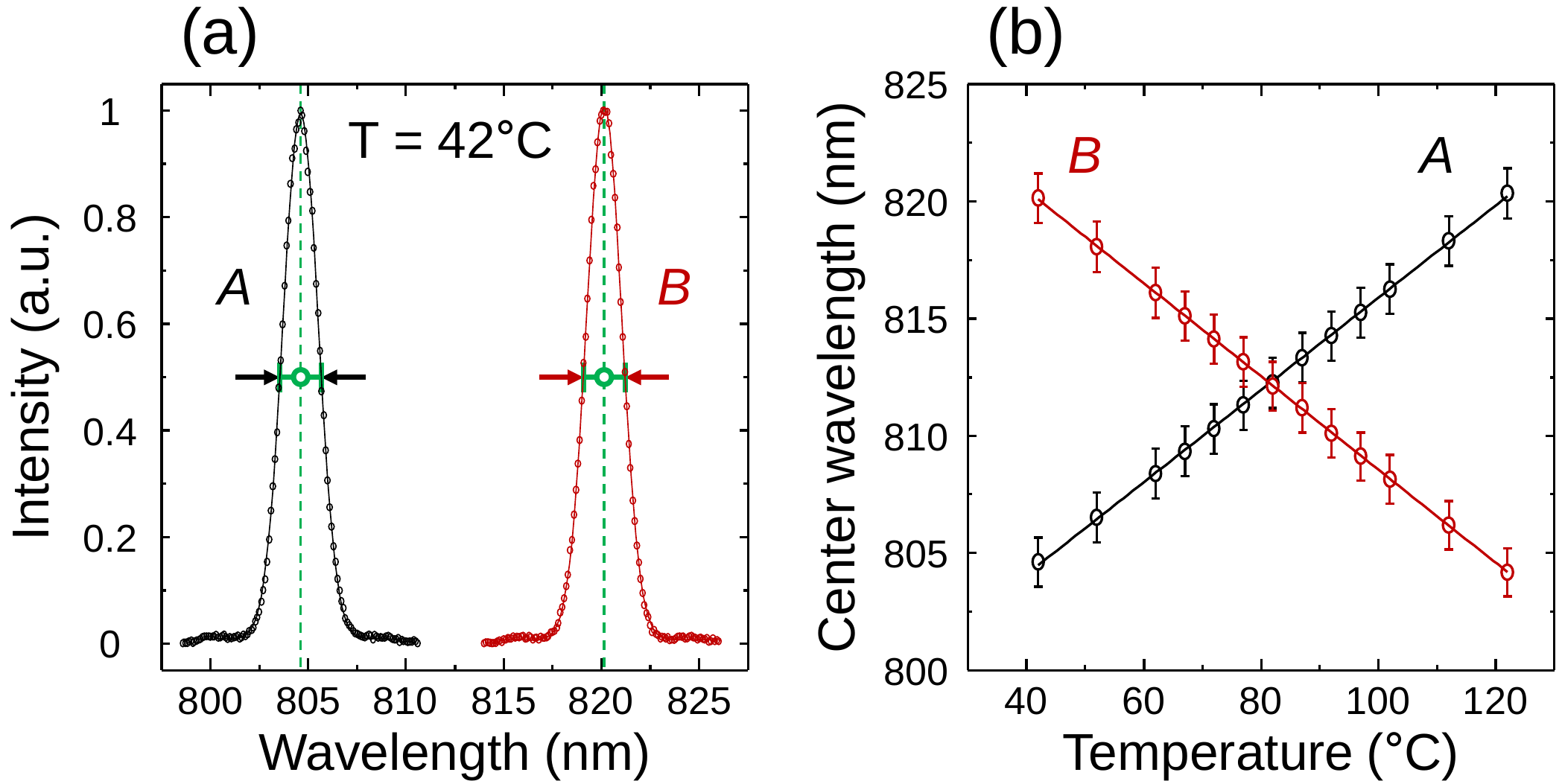}}
\caption{(a) Measured spectral distributions for photons in modes $A$ and $B$ obtained via TF at $T$ = 42$^\circ$C. (b) Center wavelength $T$ dependence for photons in modes $A$ and $B$. FWHM $\Delta \lambda$ are represented by error bars. o: measured or estimated data. $-$: Gaussian or linear fitted lines.}
\label{tuneexp}
\end{figure}

Figure \ref{tuneexp} shows (a) an example of measured spectral distributions for photons in modes $A$ and $B$, which were obtained using the TF at $T$ = 42$^\circ$C, and (b) $\lambda_{H,V}$ and $\Delta \lambda$ (represented by error bars) estimated from measured photon spectra at $T$ = 42-122$^\circ$C. The error bars in Fig.~\ref{tuneexp}(b) represent the FWHMs of the Gaussian fitted curves of the data shown in Fig.~\ref{tuneexp}(a). From the linear fitted lines of the estimated center wavelengths in Fig.~\ref{tuneexp}(b), the experimental wavelength tunability via control of $T$ was estimated as 0.198(1) nm/$^\circ$C, and $T_{dc}$ was obtained as 81.7$^\circ$. The average value of the estimated $\Delta \lambda$ (FWHM) of the photons in modes $A$ and $B$ was 2.11(3) nm. This is significantly larger than the numerically calculated values ($<$ 0.6 nm), because of the broad TF bandwidth ($\Delta \lambda$ = 1.82 nm). For the reason, we could not measure the photon $\Delta \lambda$ variation in response to the crystal $T$ directly, as shown in Fig.~\ref{tune}(b).

\begin{figure}[b]
\centerline{\includegraphics[scale=0.38]{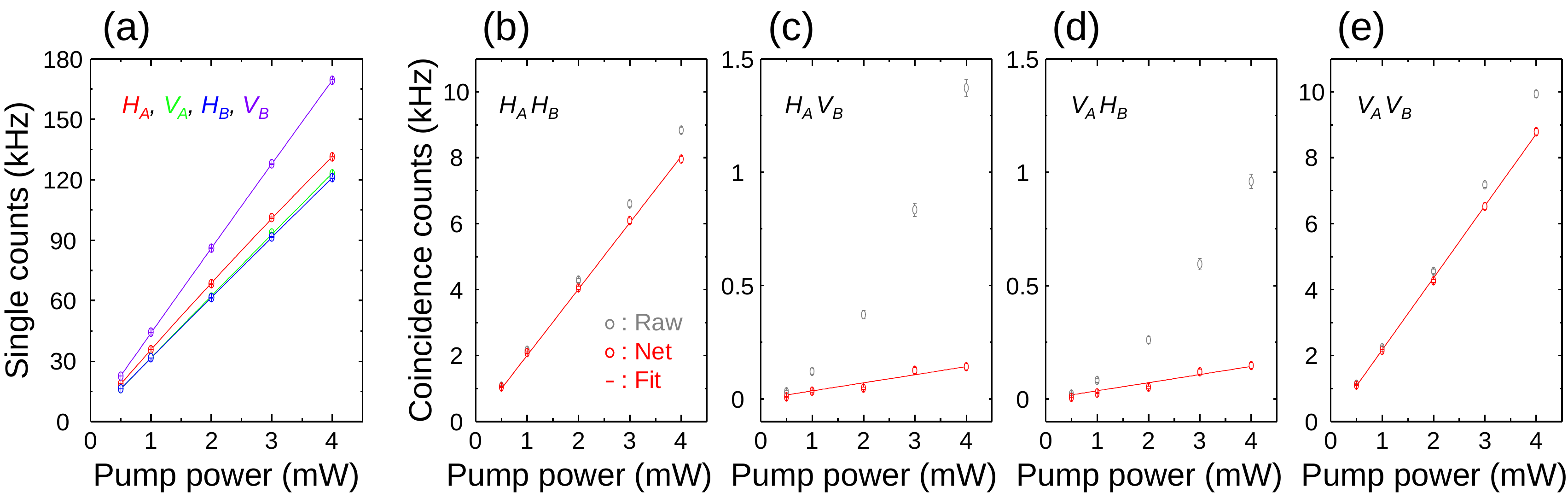}}
\caption{Pump-power $P_p$ dependence of count rates for (a) single counts of modes $H_A$, $V_A$, $H_B$ and $V_B$, and (b-e) coincidence counts of modes $H_A H_B$, $H_A V_B$, $V_A H_B$ and $V_A V_B$. Raw: Raw data of coincidence counts. Net: Accidental counts removed data. Fit: Linear fitted line of Net.}
\label{pumpcounts}
\end{figure}

\subsection{Pump power dependence of single and coincidence counts}

We measured the polarization-dependent single and coincidence count rates for modes $A$ and $B$ in Fig.~\ref{setup} with $P_p$ of 0.5-4 mW at $T$ = 82$^\circ$C around the degenerate condition. As shown in Fig.~\ref{pumpcounts}, the single and coincidence count rates were linearly proportional to $P_p$. The estimated single count rates for each mode were approximately 31-35 kHz/mW (except the result of 42.7 kHz/mW for $V_B$), whereas the coincidence count rates for modes $H_A H_B$ and $V_A V_B$ were 2.0 and 2.2 kHz/mW, respectively. The coincidence count rates for the unexpected modes $H_A V_B$ and $V_A H_B$ were approximately 36 Hz/mW, because of the finite extinction ratio of the PBS. The detected $BR$ of 4.2 kHz/mW mainly from modes $H_A H_B$ and $V_A V_B$ include several losses due to the finite transmittance ($\simeq$ 0.8) of the LP ({\it{Thorlabs}} LPVIS050), the detection efficiency ($\simeq$ 0.4 at around 810 nm) of the SPCM ({\it{PerkinElmer}} SPCM-AQ4C), and mode coupling inefficiency to the SMF, for example. Therefore, the total pair-production-rate of our source under ideal conditions is approximately 41 kHz/mW at minimum. Using the numerical result of the photon bandwidth $\Delta \lambda_{H,V}$ = 0.553 nm, the estimated spectral-pair-production-rate (pair-production-rate per bandwidth) is approximately 74 kHz/mW/nm. If we use a longer crystal with $L$ = 25 mm and scaling factor $L\sqrt{L}$~\cite{Fedrizzi}, the expected spectral-pair-production-rate is approximately 293 kHz/mW/nm. We note and expect that greater count rates can be achieved if spatial-mode optimization for the pump and photons~\cite{Fedrizzi} is implemented.

Very recently, Jeong {\it et al}. reported a bright source of polarization-entangled photons via SPDC process of type-II non-collinear PPKTP crystal and broadband multi-mode cw laser~\cite{Jeong2}. The configurations of ours and \cite{Jeong2} are the same except longitudinal modes of the pump. Jeong {\it et al}. showed that detected $BR$ using MMF collection are approximately 13 times larger than that of SMF at the cost of a little amount of concurrence. Thus, we expect that approximately one order of magnitude improvement in count rates can be achieved by MMF collection, without longer crystal and optimization of focusing parameters.

\subsection{Entanglement quality}

\begin{table}[h]
\centering\caption{Estimated entanglement qualities for various $P_p$ }
\begin{tabular}{c|c||c|c|c}
\hline
Pump power & time duration & Visibility of HOM & Fidelity with
$|\Phi^+\rangle$ & Concurrence \\
\hline
0.5 mW & 8 s & 0.974(4) & 0.981 & 0.975 \\
1 mW & 4 s &  0.972(5) & 0.980 & 0.965 \\
2 mW & 2 s &  0.957(6) & 0.978 & 0.954 \\
3 mW & 1 s & 0.961(8)  & 0.976 & 0.949 \\
4 mW & 1 s & 0.966(8)  & 0.976 & 0.956 \\
\hline
\end{tabular}
\label{pumpqualitytable}
\end{table}

\begin{figure}[h]
\centerline{\includegraphics[scale=0.45]{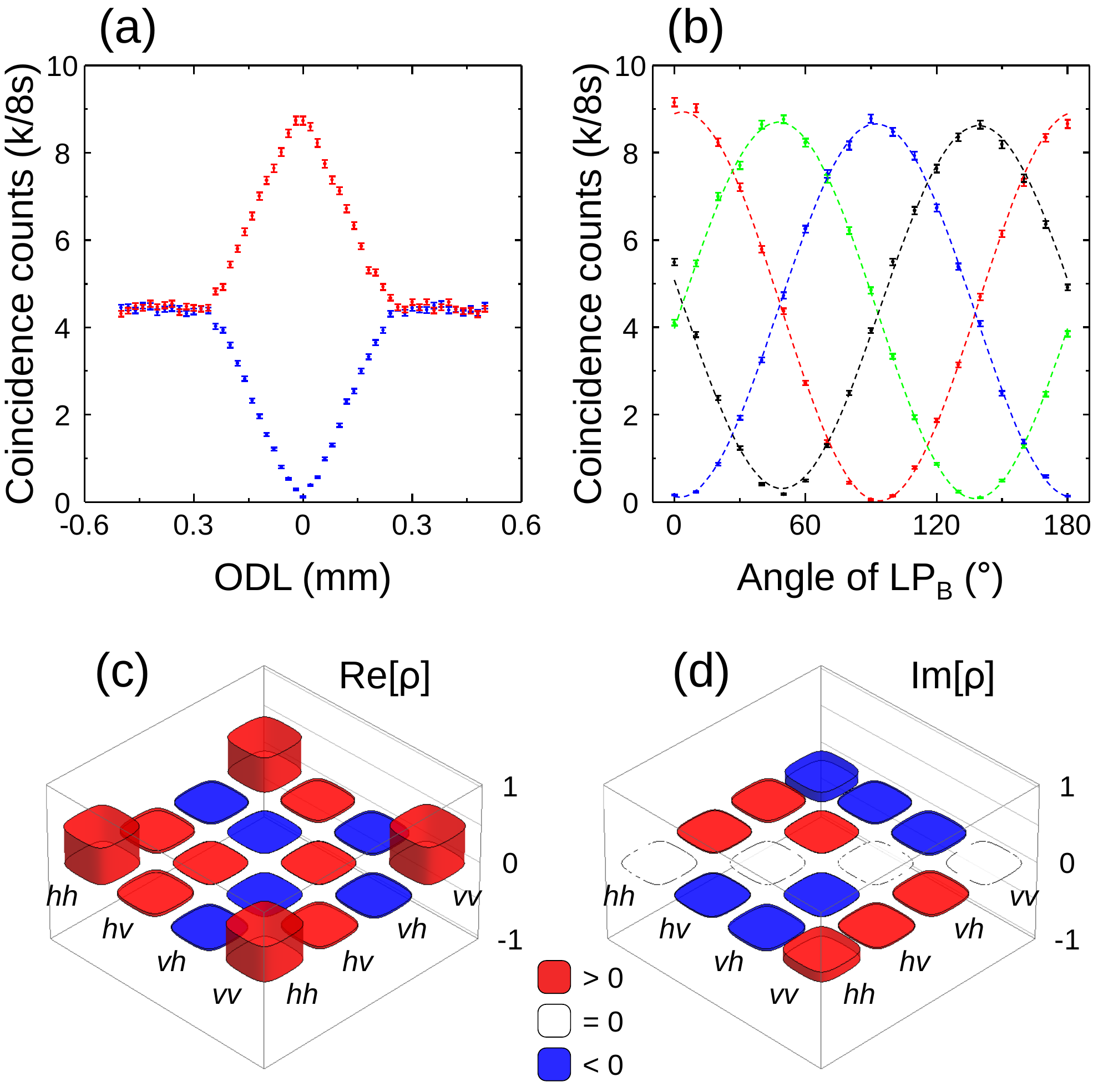}}
\caption{Experimental results at $P_p$ = 0.5 mW for: (a) HOM dip (blue) and peak (red) interferences; (b) correlation functions of polarizations for four cases; and (c) real and (d) imaginary parts of reconstructed density matrix obtained via QST. For the correlation function measurements in (b), the LP from mode $A$ (LP$_A$) was fixed to 0$^\circ$ (red); 45$^\circ$ (green); 90$^\circ$ (blue); and 135$^\circ$ (black). Dotted lines: Sine-fitted curves.}
\label{pumpquality}
\end{figure}

We estimated the entanglement quality of the generated states using several indicators: (i) The visibility of the Hong-Ou-Mandel (HOM) interferences; (ii) the correlation function visibilities; (iii) the fidelity with the ideal state ($| \Phi ^+ \rangle$) in Eq.~(\ref{phi}) for $\phi$ = 0; and (iv) the concurrence. We first investigated the $P_p$ dependence of the entanglement quality. To ensure fairness, we attempted to equalize the total number of resources by adjusting the counting duration as follows: 1, 2, 4, and 8 s for 4(3), 2, 1, and 0.5 mW, respectively. The measured quantities of the entanglement indicators for various $P_p$ are summarized in Table \ref{pumpqualitytable}, apart from the correlation function visibilities, which were approximately 0.92(1)-0.99(1). As an example, Fig. \ref{pumpquality} shows measured results for: (a) the HOM dip and peak interferences; (b) the polarization correlation functions; and (c, d) the reconstructed density matrix obtained via QST at $P_p$ = 0.5 mW.

\begin{figure}[b]
\centerline{\includegraphics[scale=0.426]{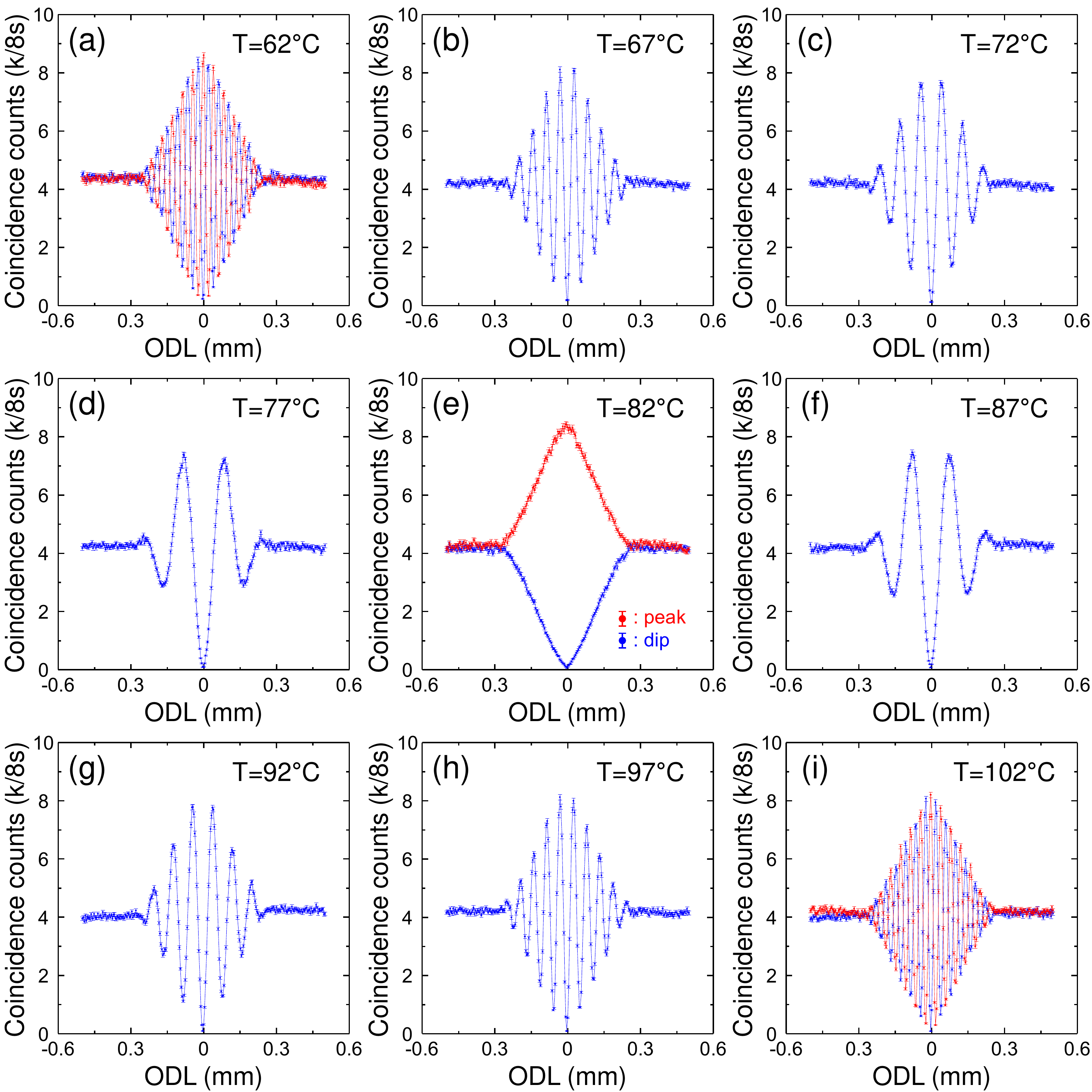}}
\caption{Measured HOM dip and peak interferences for $T$ = 62-102$^\circ$C at $P_p$ = 0.5 mW and $t$ = 8 s. Peak (dip) data in red (blue) are measured based on coincidence counts of modes $A$ and $B$ with LP$_A$ and LP$_B$ angles set to 45$^\circ$ and (-)45$^\circ$, respectively. To show the beat patterns clearly, lines are drawn between neighboring points.}
\label{HOMT}
\end{figure}

\begin{table}[t]
\centering\caption{Entanglement qualities of generated states for various $T$}
\begin{tabular}{c||c|c|c|c|c|c|c|c|c}
\hline
$T$&{62$^\circ$C}&{67$^\circ$C}&{72$^\circ$C}&{77$^\circ$C}&{82$^\circ$C}&{87$^\circ$C}&{92$^\circ$C}&{97$^\circ$C}&{102$^\circ$C}\\
\hline
Fidelity&{0.96}&{0.97}&{0.97}&{0.98}&{0.98}&{0.98}&{0.98}&{0.98}&{0.98}\\\hline
Concurrence&{0.97}&{0.98}&{0.98}&{0.97}&{0.97}&{0.97}&{0.97}&{0.97}&{0.97}\\\hline
\end{tabular}
\label{Tqualitytable}
\end{table}

For the HOM interference and polarization correlation measurements, the QWPs for modes $A$ and $B$ in Fig.~\ref{setup} were fixed to 0$^\circ$ and 90$^\circ$, respectively, so as to compensate for the phases introduced by the QWPs themselves. The HOM peak (dip) interference was measured using the coincidence counts between modes $A$ and $B$ by sweeping the position of the right-angled prism (ODL) in Fig.~\ref{setup}. The LPs for modes $A$ and $B$ (LP$_A$ and LP$_B$, respectively) were set to 45$^\circ$ and (-)45$^\circ$, respectively. We also measured the correlation functions of the polarizations in modes $A$ and $B$ for four LP$_A$ values of 0$^\circ$, 45$^\circ$, 90$^\circ$, and 135$^\circ$, the results of which were fit using raised sine functions. The visibilities obtained from the fitting results were 0.994, 0.982, 0.975 and 0.930, respectively. To calculate the fidelities with the ideal state and concurrences, we performed 2-qubit QST and reconstructed the density matrices. As shown in Table \ref{pumpqualitytable}, when $P_p$ increases the entanglement qualities decrease, as the multi-photon effects are increased and the coincidence window is relatively large. We expect that higher visibility and concurrence ($>$ 0.97) can be measured with a shorter coincidence window at lower $P_p$.

To investigate the entanglement qualities for non-degenerate cases, we also measured the quality indicators for varying $T$. Figure \ref{HOMT} shows that HOM dip (in blue) and peak (in red) interferences at $T$ = 62-102$^\circ$C, $P_p$ = 0.5 mW, and $t$ = 8 s. The plotted points are net coincidence counts for modes $A$ and $B$. All the visibilities of the interference patterns in Fig.~\ref{HOMT} are larger than 0.97. The fidelities and concurrences estimated from the QST results are summarized in Table~\ref{Tqualitytable}. The entanglement qualities for non-degenerate cases ($\left| \lambda_{A} - \lambda_{B} \right | <$ 16 nm) are almost identical in our setup, as we used zero-order wave plates for 808 nm and broadband PBS and LP to reduce wavelength-dependent systematic errors.

\subsection{Phase modulation and stability}

\begin{figure}[h]
\centerline{\includegraphics[scale=0.5]{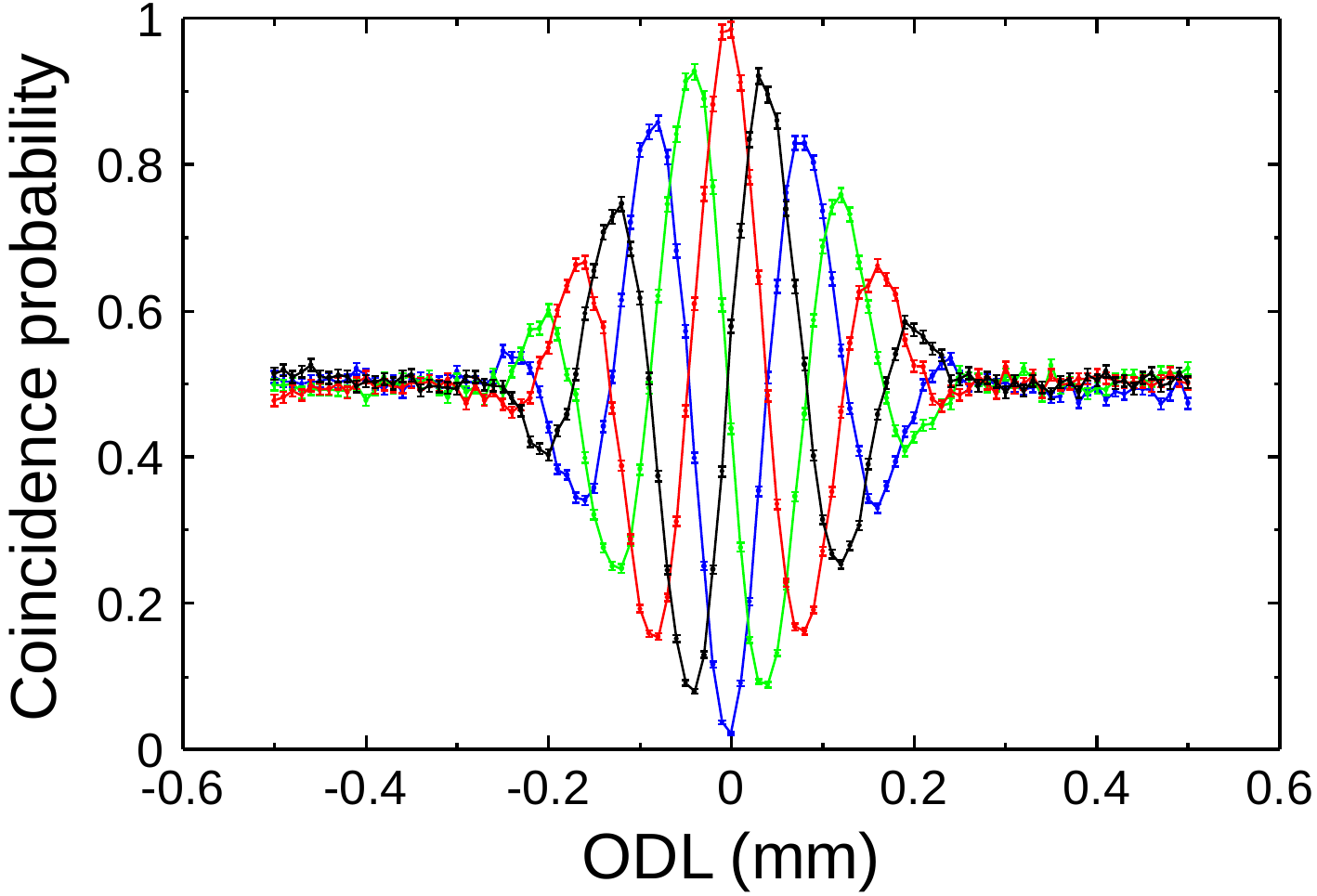}}
\caption{HOM dip interferences of generated state, $|\lambda_A \lambda_B \rangle_{AB} \otimes  \left(|HH\rangle_{AB} + e^{i\phi} |VV \rangle_{AB} \right) /\sqrt{2}$, where $\phi$ = $n \pi / 2$ and $\lambda_A \neq \lambda_B$. The datapoints colored in blue, green, red, and black correspond to $n$ = 0, 1, 2, and 3, respectively.}
\label{HOMP}
\end{figure}

The $\phi$ between $|HH\rangle_{AB}$ and $|VV\rangle_{AB}$ in Eq.~(\ref{phi}) can be modulated using the phase shifter shown in Fig.~\ref{setup}. To check the phase modulability, we measured the HOM dip interferences for the cases of $\phi = n \pi / 2$, where $n$ = 0, 1, 2 and 3. Figure \ref{HOMP} shows the normalized HOM dip (LP$_A$ = 45$^\circ$ and LP$_B$ = -45$^\circ$) beating patterns at $P_p$ = 0.5 mW, $t$ = 8 s, and $T$ = 77$^\circ$C. The datapoints colored blue, green, red, and black correspond to $n$ = 0, 1, 2, and 3, respectively. The total duration of the data measurement for the results shown in Fig.~\ref{HOMP} was more than 1 h.

The experimental configuration, especially the compensation section, employed in the approach presented in this study is based on MZI. Thus, the phase stability is fundamentally limited, because of the vibrations of the optic mounts. However, the Sagnac configuration is robust against this effect. If we denote the two optical path lengths from the PPKTP crystal to the PBS in modes $a$ and $b$ in Fig.~\ref{setup} as $l_a$ and $l_b$, respectively, the $\phi$ difference between the superposed two states in Eq. (\ref{phi}) induced by the difference in optical path length is $(l_a - l_b)(k_A - k_B)$, where $k_{A(B)}$ = $2 \pi / \lambda_{A(B)}$. Therefore, the total phase fluctuation $\Delta \phi$ due to the optics vibration $\Delta l$ in the MZI can be expressed as $\sqrt{m} \left|\delta k\right| \Delta l$, where $\delta k = k_A - k_B$  and $m$ is the number of independent optics in the MZI. The order of $\Delta l$ is typically sub-micrometer in the case on a vibration-isolated optical table. For example, for parameters of $\delta \lambda$ = $\lambda_B - \lambda_A$ $\simeq$ 50 nm, $\lambda_p = 406.2$ nm, $m$ = 5, and $\Delta l$ = 0.1 $\mu$m, the expected value of $\Delta \phi$ is approximately 0.017 $\times$ 2$\pi$ ($<$ 2\%). Thus, the phase fluctuations are negligible ($< 0.4\%$) in our experimental results ($\delta \lambda <$ 10 nm). Moreover, for the degenerate case ($\lambda_A$ = $\lambda_B$), $\Delta \phi$ can be fully neglected (as in the Sagnac configuration), because $\delta k$ = 0. This may be comprehensive in terms of quantum mechanics. The two superposed states $|HV\rangle_{ab}$ and $|VH\rangle_{ab}$ of a photon pair immediately after generation via SPDC are simultaneously influenced by the phase fluctuations of the optics in the MZI; thus, this phase can be factored out as an overall phase.

\section{Summary} \label{s5}
We have demonstrated a stable and bright polarization-entangled photon-pair source obtained via a type-II non-collinear SPDC process with a 10-mm-long PPKTP crystal. The detected brightness was 4.2 kHz/mW, and the spectral-pair-production-rate was estimated as being 74 kHz/mW/nm. Further, we expect that a spectral-pair-production-rate of more than 293 kHz/mW/nm can be achieved via optimization of the focusing conditions and through use of a 25-mm crystal, or photon collection using MMF as previous studies ~\cite{Fedrizzi, Jeong2}. This expectation is based on the fact that the non-collinear setup can collect twice the number of photon pairs as a collinear setup. Our setup is as stable as the Sagnac configuration for the degenerate case. Although this stability decreases with increased wavelength non-degeneracy, the instability is acceptable if the non-degeneracy is not excessive compared with the MZI compensator instability. For various photon-pair wavelengths ($\lambda_{A,B}$: 808-816 nm) obtained by controlling the crystal temperature (62-102$^\circ$), the fidelities with one of the Bell states ($|\Phi^+\rangle$) and the concurrences of the generated states were estimated as being approximately 0.97-0.98, despite the large coincidence window (55 ns). Therefore, our source is bright, wavelength-tunable, and relatively stable (comparable to the Sagnac configuration setup) with no requirement for specialized optical components. Further, this source has high entanglement quality; thus, we expect that it will be utilized in various experiments concerning quantum information processing.

\section*{Acknowledgments}
This work was supported by the National Research Foundation of Korea (NRF) grant funded by the Korea government (MSIP) (No. 2015R1A2A1A05001819, No. 2014R1A1A2054719, and No. 2014R1A1A2055488), and by the Measurement Research Center (MRC) Program for Korea Research Institute of Standards and Science.


\begin{thebibliography}{99}
\bibitem{elliptic}
The authors of \cite{Fiorentino} have claimed that the slightly elliptic ring image obtained under the noncollinear degenerate condition stems from the biaxiality of the KTP crystal. However, this is ambiguous, as elliptic ring images can be generated from uniaxial crystals ($n_x = n_y \neq n_z$). In addition, KTP is approximately uniaxial ($n_x \simeq n_y < n_z$).
\end{thebibliography}
\end{document}